\documentclass[aps,prl,twocolumn,showpacs,superscriptaddress]{revtex4-1}
\usepackage{graphicx}
\usepackage{amssymb}
\usepackage{amsmath}
\usepackage{mathtools}
\usepackage[utf8]{inputenc} 
\usepackage[T1]{fontenc}

\begin{document}

\title{Permanent mitigation of loss in ultrathin SOI high-Q resonators using UV light}

\author{Gioele Piccoli}
\affiliation{Centre for Materials and Microsystems, Fondazione Bruno Kessler, I-38123 Povo, Italy}
\affiliation{Department of Physics, Nanoscience Laboratory, University of Trento, I-38123 Povo, Italy}

\author{Martino Bernard}
\affiliation{Department of Information Engineering, University of Brescia, I-25123 Brescia, Italy}	

\author{Mher Ghulinyan}
\affiliation{Centre for Materials and Microsystems, Fondazione Bruno Kessler, I-38123 Povo, Italy}

\date{\today}

\begin{abstract}
Mitigation of optical losses is of prime importance for the performance of integrated micro-photonic devices. In this paper, we demonstrate strip-loaded guiding optical components realized on a 27~nm ultra-thin SOI platform. The absence of physically etched boundaries within the guiding core suppresses majorly the scattering loss, as shown by us previously for a silicon nitride (Si$_3$N$_4$) platform [Stefan \textit{et. al.}, OL 40, 3316 (2015)]. Unexpectedly, the freshly fabricated Si devices showed large losses of 5 dB/cm, originating from absorption by free carriers, accumulated under the positively charged Si$_3$N$_4$ loading layer. We use 254 nm ultraviolet (UV) light exposures to neutralize progressively and permanently silicon nitride’s bulk charge associated with diamagnetic K+ defects. This in turn leads to a net decrease of electron concentration in the SOI layer, reducing thus the propagation loss down to 0.9 dB/cm. Detailed MOS-capacitance measurements on test samples were performed to monitor the UV-induced modification of the electronic properties of the system. The evolution of loss mitigation was directly monitored both by Beer-Lambert approach in waveguide transmission experiments, as well as through more accurate cavity linewidth measurements. In the last case, we demonstrate how intrinsic cavity Q’s boost from 60,0000 to up to 500,000 after UV treatment. Our results may open routes towards engineering of new functionalities in photonic devices employing UV-modification of space charges and associated local electric fields, unveil the origin of induced optical nonlinearities in Si$_3$N$_4$/Si micro-photonic systems, as well as envisage possible integration of these with ultra-thin SOI electronics.
\end{abstract}

\pacs{}
	
\maketitle
\section{Introduction}

Light confinement by resonant circulation of electromagnetic radiation in mirrorless microresonators makes them a key building block for planar integrated photonics \cite{rabus2007integrated,heebner2008optical}. The amount of optical power which is lost per round trip of circulation manifests in the spectral width of resonances: the smaller is the loss the narrower resonances get. In general, the possible channels of \textit{intrinsic} loss $\alpha_i$ in a cavity are the material absorption, $\alpha_{m}$, the scattering due to boundary imperfections, $\alpha_{sc}$, and out-radiation, $\alpha_{rad}$, due to the curved geometry. The overall intrinsic loss of a high-finesse cavity, combined with the \textit{extrinsic} loss $\alpha_{e}$ due to coupling to an external waveguide, thus define the spectral width of a Lorentzian-shaped resonance peaked at a frequency of $\omega_0$ via
\begin{equation}
f(\omega)\sim \frac{\alpha_{e}}{(\alpha_i+\alpha_{e})/2-\imath (\omega-\omega_0)}.
\label{eq:lor}
\end{equation} 
A reduced intrinsic loss is essential for a number of applications such as passive filtering in optical communication networks 
\cite{dai2012passive,subramanian2013low,zhuang2011low}, quantum optics \cite{engin2013photon,grassani2015micrometer,silverstone2015qubit,silverstone2016silicon,christensen2017engineering}, space \cite{kippenberg2011microresonator,pasquazi2013self} or sensing \cite{armani2006heavy,zhu2010chip,samusenko2016sion}. Suppressing the intrinsic loss to only that of the material itself is challenging for it can push the device characteristics to an ultimate limit \cite{lee2012chemically}. Typical approaches to achieve minute contributions from radiative and scattering loss are strong modal confinement (high-index contrast and large radius of curvature) \cite{lee2012chemically,stefan2015ultra,nezhad2011etch,ji2017ultra} and realization of smooth device boundaries during fabrication \cite{nezhad2011etch,griffith2012high,stefan2015ultra,lee2012chemically}, respectively.

In an optical device, the material loss $\alpha_{m}$ is largely dependent on the choice of the operation frequency. In particular, most silicon micro-photonic devices operate at telecom frequencies, where the photon energy is below silicon's electronic band gap \cite{soref2006past,vivien2016handbook}. The interband absorption, thus, is negligible, and the associated loss can be as low as $0.005$~dB/cm for a typical p-type Si of 15 $\Omega\cdot$cm resistivity. On the other hand, a non-negligible intraband two-photon absorption (TPA) and excited-carrier absorption (ECA) can increase the loss by up to three orders of magnitude \cite{liang2004role,priem2005optical,leuthold2010nonlinear}. These last can be still mitigated by integrating p-n junction devices to operate under reverse-bias conditions, depleteing thus silicon from electrical charge carriers in the guiding region \cite{vivien2016handbook}.

Silicon nitride is widely used in integrated circuit technology \cite{milek2013silicon}, flat-panel displays \cite{kanicki1992amorphous} and solar cells \cite{schropp1998amorphous}. Since a decade, silicon nitride is attracting the photonics community for CMOS-compatibile integrated photonics. It is a key dielectric material for linear micro-optical guiding circuits, transparent from visible to MIR wavelengths \cite{gorin2008fabrication,subramanian2013low,romero2013silicon}, as well as for applications in nonlinear frequency conversion schemes either due to it's intrinsic nonlinearities \cite{levy2011harmonic,okawachi2011octave,moss2013new} or induced nonlinearities when a silicon nitride film is applied to crystalline silicon \cite{jacobsen2006strained,cazzanelli2012second,schriever2015second}.

\begin{figure*}[t!]
	\centering
\includegraphics[width=1.8\columnwidth]{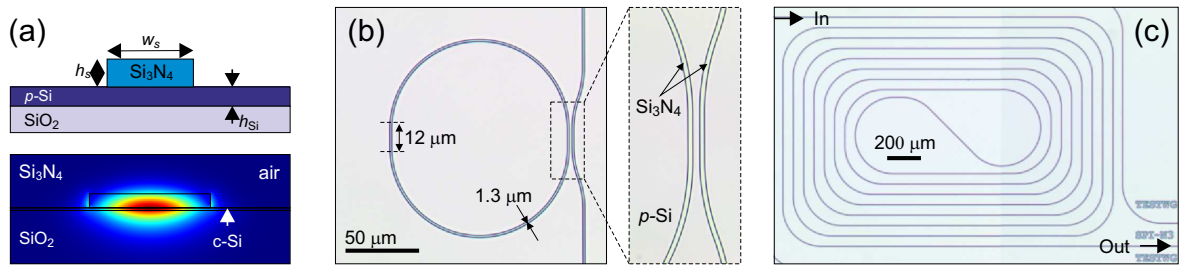}
\caption{(a) Top - A cross-sectional schematics of the nitride-loaded ultrathin SOI device. Bottom - The numerical FEM calculation of fundamental TE mode intensity profile at a wavelength of 1550~nm. (b) A top-view optical micrograph of a ring resonator device. A blow-up of the image on right shows the optical coupling region. (c) An optical micrograph of a spiral waveguide used for propagation loss measurements. The waveguide width is 1.3 $\mu$m and the shown spiral has a total length of 3 cm. }
\label{fig:fig1}
\end{figure*}

Here we report on the design, fabrication and characterization of high-Q micro-optical components on an ultra-thin, 27~nm thick SOI platform, where light guiding is enabled by a patterned layer of strip-loading stoichiometric Si$_3$N$_4$. The absence of physically etched device boundaries in the SOI layer is expected to provide with ultimately low losses, limiting them to that of the intrinsic absorption of the lightly doped p-type silicon device layer. Such an approach was successfully implemented in our earlier study, where $Q\sim 4\times 10^6$ were achieved on a 80~nm thick Si$_3$N$_4$ platform \cite{stefan2015ultra}. Surprisingly, the freshly fabricated silicon devices showed unexpectedly high propagation losses of up to 5 dB/cm. We related the origin of these losses to the absorption of free carriers within the Si layer, which are generated due to the presence of fixed positive charges in the deposited silicon nitride. Next, we neutralized successfully and permanently the charge in Si$_3$N$_4$ by exposing the devices to 254~nm wavelength UV light. This led to an improvement of losses down to 0.9 dB/cm, allowing to boost the quality factors of ring resonator devices from an initial 60,000 up to 500,000. 

Our results open the door to the implementation of UV-induced charge modification for the design and study of new photonic devices, where the local static electric field can be engineered such to realize integrated devices with modulated linear and nonlinear optical characteristics, or, even, to cancel permanently space charge-induced fields. We also envisage the possibility of compact integration of micro-photonic components with MOFSET electronics on the same ultra-thin SOI platform in the feature.

\begin{figure*}[t]
	\centering
\includegraphics[width=1.96\columnwidth]{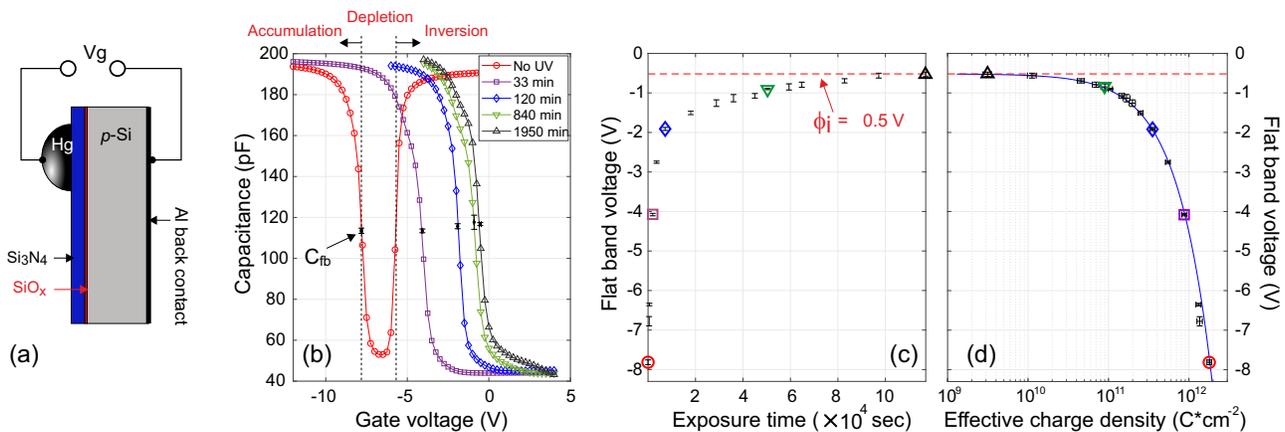}
\caption{(a)  A cross-sectional schematics of the MOS device for C-V measurements. A 145~nm Si$_3$N$_4$ layer was deposited on top of a p-type silicon substrate with a resistivity of 15~$\Omega\cdot$cm. A thin 5~nm SiO$_x$ layer, grown during the RCA clean, is present between the Si$_3$N$_4$ film and the substrate. The gate contact is formed by a Hg droplet of 787~$\mu$m diameter. (b) Selected low-frequency (10~kHz, quasi-static) C-V curves, measured after UV exposures of different duration. The C-V-response of the reference (UV-untreated) device indicates to a presence of a net positive charge and is in the conditions of strong \textit{inversion} at 0~V bias. (c) The extracted flat-band voltage as a function of UV exposure time shows a monotonic shift towards lower voltages, which is an indication of significant charge neutralization. The shift of $V_{fb}$ saturates approaching the metal-semiconductor work function potential at $\phi_{i}\sim -0.52$~V. (d) The corresponding charge density variation shows a three orders of magnitude decrease with respect to the initial situation. The solid line is a linear fit to $\sigma(V_{fb})$ with an absolute slope value of $2.5\times10^{11}$~cm$^{-2}$V$^{-1}$.}
\label{fig:CV}
\end{figure*}

\section{Materials and methods}
\label{sec:matmet}
\subsection{Device fabrication}
\label{fab}
The samples were realized starting from 6" SOI wafers with a 3~$\mu$m-thick buried oxide (BOX) and a 250~nm (100)-Si device layer (Soitec). The latter is lightly-doped with Boron and has a nominal resistivity of 15 $\Omega\cdot$cm. First, the device layer thickness was reduced by thermally oxidizing the Si and removing the grown oxide in a buffered HF solution. A fine tuning of the final thickness, $29.7\pm1.5$ nm was achieved by means of standard RCA cleaning steps \cite{kern1970cleaning}. Next, a 145 nm thick LPCVD Si$_3$N$_4$ film was deposited at 780$^\circ$C and patterned lithographically using an i-line Nikon stepper. The pattern was then transferred to Si$_3$N$_4$ using an inductively-coupled plasma etch, terminating with a wet etching step for last 10~nm's to guarantee a smooth top surface of the underlying Si layer. Finally, the fabricated chips were diced using a polishing-grade saw to define the waveguides input-output facets. The cross sectional view of a generic device and the distribution of the simulated electric field intensity of a Transverse Electric (TE) polarized mode are shown in Fig.~\ref{fig:fig1}(a). Optical micrographs of a typical ring resonator and a spiral waveguide are shown in Figs. ~\ref{fig:fig1}(b) and (c), respectively. 

\subsection{Optical characterization}
The devices were tested in waveguide transmission experiments in a 100 nm range around 1550 nm of wavelength. A tunable laser (Yenista Tunics T100S HP) was butt-coupled to the waveguides using a lensed fiber (total insertion loss of $3.8\pm0.2$ dB/cm). The signal polarization was controlled at the waveguide input, while the transmitted signal was collected with a second fiber at the waveguide output. The signal was then sent to an InGaAs photodiode and analyzed using a high resolution oscilloscope (PicoScope 4224).

\subsection{MOS capacitance measurements}
Quasi-static C-V measurements were conducted at 10~kHz using an MDC 802-150 Mercury Probe (spot diameter 787~$\mu$m) and acquired with an Agilent 4156C Parameter Analyzer. Different samples, cut from the same wafer, were exposed to 254~nm UV light from a Hg bulb lamp (254~nm, 19~mW/cm$^2$). The exposure times were varied from 1 minute up to 32 hours. As a reference, we measured also an as-deposited sample with no UV exposure. In order to acquire enough statistics per sample, the C-V measurements were repeated on at least five different points on each sample. 

Based on the results from C-V experiments, several chips, containing microphotonic components, such as cm-long spirals and ring resonators, were exposed to UV light in order to study directly the variation of optical characteristics of these devices due to electrical charge neutralization.

\section{Results and discussion}
\subsection{Diamagnetic and paramagnetic centers in silicon nitride}
\label{sec:MOS}
Silicon nitride, both in its stoichiometric (Si$_3$N$_4$) and N-rich form ($\alpha-$SiN$_x$:H), is known to host charge trapping centers via silicon dangling bond defects. Such defects are typically present and homogeneously distributed in the film either in a neutrally charged paramagnetic state, known as the $K^0$-centers ($\cdot$Si$\equiv$N$_3$), or in a charged diamagnetic state \cite{krick1988electrically}. These last can be positively ($K^+$, $^+$Si$\equiv$N$_3$) or negatively charged ($K^-$, $^-$Si$\equiv$N$_3$) \cite{kumeda1984photo,krick1988electrically,warren1990first,warren1991electrically,kobayashi2011ultraviolet}. Interestingly, exposing films to an ultraviolet (UV) radiation with energies greater than 3.5~eV ($<350$~nm) can induce a change in the spin and charge state of diamagnetic states according to a reaction \cite{warren1993electron}
\begin{equation}
K^- + K^+ +h\nu \rightarrow 2K^0,
\label{eq:K}
\end{equation}
increasing the concentration of neutral $K^0$-centers and, therefore, leading to a partial or complete compensation of space charge in the film \cite{krick1988electrically}. From a point of view of device functionality, UV-induced charge reduction can modify largely the electrical characteristics, for example, via a redistribution of space charge in the semiconductor material in contact with the film.

Capacitance measurements of a dielectric film in a two-terminal MOS configuration offer a wealth of information on the fabrication process, in particular, the nature, sign and amount of electrical charge in the bulk of the dielectric film and at the interface with the semiconductor substrate (Fig.~\ref{fig:CV}(a)). In order to reveal and quantify  possible charges within the silicon nitride, we have deposited a 145~nm thick Si$_3$N$_4$ film on top of Si substrates and performed C-V measurements in a MOS configuration. For this, Boron-doped p-type substrates with 15~$\Omega\cdot$cm resistivity, matching with that of the device layer of our SOI wafers, were chosen. In addition, prior to Si$_3$N$_4$ deposition, a thin 5nm SiO$_x$ layer was grown during a standard RCA cleaning step. Thus, the test wafers replicate exactly the strip-loaded SOI devices. 

We found that the as-deposited nitride layer contains a large amount of positive electrical charge. The typical C-V curve for this reference sample (Fig.~\ref{fig:CV}(b), circles) passes from an "accumulation" to a "depletion" state at negative gate voltages, with a characteristic flat band voltage value of $V_{fb}\approx-7.8$~V. This last is estimated graphically, once the flat band capacitance $C_{fb}$ is calculated following
\begin{equation}
C_{fb}=\frac{C_{max}\epsilon_{s}\epsilon_0 A/L_D}{C_{max}+\epsilon_{s}\epsilon_0 A/L_D},
\label{eq:Cfb}
\end{equation}
where $C_{max}$ is the film capacitance in accumulation regime, $\epsilon_{s}=11.68$ and $\epsilon_{0}$ are the substrate and the vacuum permittivities, respectively, $A$ is the area of the Hg droplet contact and $L_D\sim180$~nm is the extrinsic Debye length in the Si substrate. Below $V_g\sim-6$~V the capacitance grows again, indicating to an "inversion" of the conductivity type at the surface of the substrate from p- to n-type. The areal density of the corresponding net positive charge amounts to $\sigma=1.7(\pm0.1)\times 10^{12}$~cm$^{-2}$, which is in agreement with previous estimations \citep{schriever2015second}. 

Capacitance measurements on UV-exposed samples show a gradual shift of the C-V curves towards positive voltages as a function of exposure time. Repeated measurements at a distance of a couple of weeks showed that this shift is permanent. Selected examples of C-V curves are shown in Fig.~\ref{fig:CV}(b), while in Fig.~\ref{fig:CV}(c) we show the dependence of the extracted flat-band voltage $V_{fb}$ on UV illumination time. We notice that $V_{fb}$ decreases exponentially within the first couple of hours of exposure. For the next 30 hours of exposure $V_{fb}$ continues to shift monotonically at a much slower rate (almost linearly) towards the asymptotic value of metal-semiconductor work function $\phi_{i}$, which for our Hg/p-type Si system amounts to $-0.52$~V. 

The corresponding variation of charge areal density $\sigma$ against $V_{fb}$ is shown in Fig.~\ref{fig:CV}(d). The UV illumination decreases the positive charge density by three orders of magnitude, leading to a change of the flat-band voltage from it's initial value down to $\phi_{i}$ for the longest exposures. The estimated residual charge density amounts to $\sigma=3.1\times10^9$~cm$^{-2}$, which is comparable to the density of charge traps of very-high-quality oxide/Si interfaces or the density of dopant ions per cm$^2$ for a silicon layer of $15~\Omega\cdot$cm resistivity \cite{zhang2006electronic}.

\subsection{UV exposure effect on optical losses in microphotonic devices}
Following the encouraging results obtained from MOS capacitance measurements, we studied the evolution of optical losses of Si$_3$N$_4$-loaded SOI devices in response to UV light illumination. For this, several chips, containing both spiral waveguides of different length and ring resonators, where characterized in optical transmission experiments prior and after UV exposure.

\textit{Loss characterization from waveguides} -- The studied waveguides were composed of 27~nm-thick continuous SOI slab and a 145~nm-thick Si$_3$N$_4$ loading strip of 1300 nm width. Finite-element numerical simulations, performed in the design phase, suggest that this geometry does not guide the Transverse Magnetic (TM) polarization and supports one single TE mode in the 1.5-1.6~$\mu$m wavelength range, while the Si slab alone does not guide light. 

\begin{figure}[t!]
	\centering
\includegraphics[width=0.95\columnwidth]{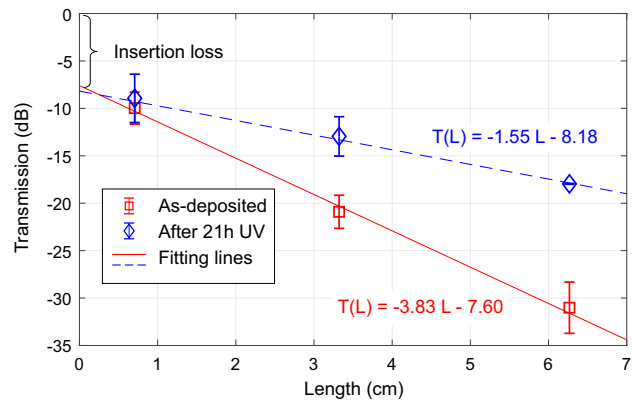}
\caption{The attenuation of propagating optical power was measured for waveguides of different length prior (red, $\square$) and after UV exposure for 21 h (blue, $\diamond$). The error bars represent the statistical error over similar devices. A Lambert-Beer fit (lines) to the experimental data reveals a net improvement of the propagation loss due to reduced free-carrier absorption as a result of neutralization of positive charge in Si$_3$N$_4$. Note that the UV treatment does not affect the insertion loss of waveguides. }
\label{fig:LB}
\end{figure}

Figure \ref{fig:LB} reports the results of waveguide transmission experiments. The measured waveguides had lengths of 0.61 cm, 3.22 cm and 6.17 cm. The propagation losses in as-prepared devices (red, $\square$) amount to $3.83 \pm0.26$ dB/cm according to Beer-Lambert law \cite{beer1852determination}, and, if attributed to sidewall scattering, are unexpectedly high for the considered strip-loaded configuration and the adopted processing technology \cite{stefan2015ultra}. As anticipated in the section \ref{sec:MOS}, we expect that the net positive charge in the loading Si$_3$N$_4$ layer recalls negative charge carriers in the underlying Si core, increasing thus the free-carrier absorption within the waveguide. In fact, the same devices show much improved characteristics after 21 h of UV exposure (blue, $\diamond$), and the resulting propagation loss is decreased down to $1.55 \pm 0.28$ dB/cm. 

According to Figs.~\ref{fig:CV}(c),(d) the areal charge density is estimated to decrease from $2 \times 10^{12}$ cm$^{-2}$ in as-prepared samples down to $5 \times 10^{10}$ cm$^{-2}$ after 21 h of UV exposure. From total charge balance conditions, the equivalent free electron density, $N_e$, in the Si core would reduce from a value of $7 \times 10^{17}$ cm$^{-3}$ to $2.5 \times 10^{16}$ cm$^{-3}$. In a first order approximation, it is possible to relate the change in $N_e$ to the absorption loss $\alpha$ within a crystalline Si core by assuming that: 
\begin{itemize}
\item[$\bullet$] all losses are due to free carrier-induced absorption and no scattering losses are present,
\item[$\bullet$] no other sources of charge are present in the vicinity of the mode (e.g. charges at the Si/BOX interface),
\item[$\bullet$] the effective loss coefficient is reduced from its bulk value proportional to the power confinement factor $\Gamma$ of the mode within the Si layer ($\Gamma=0.29$ from FEM calculations at $\lambda=1.55\mu$m).
\end{itemize} 
We calculated the expected $\alpha(N_e)$ using different empirical models known from literature \cite{soref1987electrooptical,degallaix2013bulk,nedeljkovic2011free}. Such values range from 2 to 6 dB/cm for as-prepared samples and from 0.15 to 0.25 dB/cm after 21 hours of UV-exposure. It is clear, that while our estimation of $\alpha\approx 3.83$ dB/cm for the first case is well within the theoretical range, the ``UV-exposed" case with $\alpha\approx 1.55$ dB/cm is larger by more than an order of magnitude with respect to expectations.

Our conclusion at this point is, that either additional loss source is present in the studied devices, or the Beer-Lambert approach is not enough precise in the conditions where few experimental data are available and the insertion losses from one to another waveguide differ due to facets imperfections. This last is in fact reflected by the large error bars of experimental points in Fig.~\ref{fig:LB}, given by the variance over three different chips . This limitation can be surpassed by extracting the loss from spectral characteristics of ring resonators \cite{stefan2015ultra}, which are less affected by fluctuations of the waveguide facet quality. 

\textit{Loss characterization from resonators} -- A generic circular resonator induces spectral dips in the transmission spectrum of the waveguide to which it is coupled. These dips become visible as soon as the resonator's intrinsic loss, $\alpha_i$, and  the coupling loss, $\alpha_{e}$, become comparable (see Eq.~\ref{eq:lor}). In particular, for a fixed geometry, i.e. a constant $\alpha_{e}$, the resonator's spectral dips start to spot out from the bare waveguide's transmission spectrum. 

\begin{figure}[t!]
	\centering
\includegraphics[width=\columnwidth]{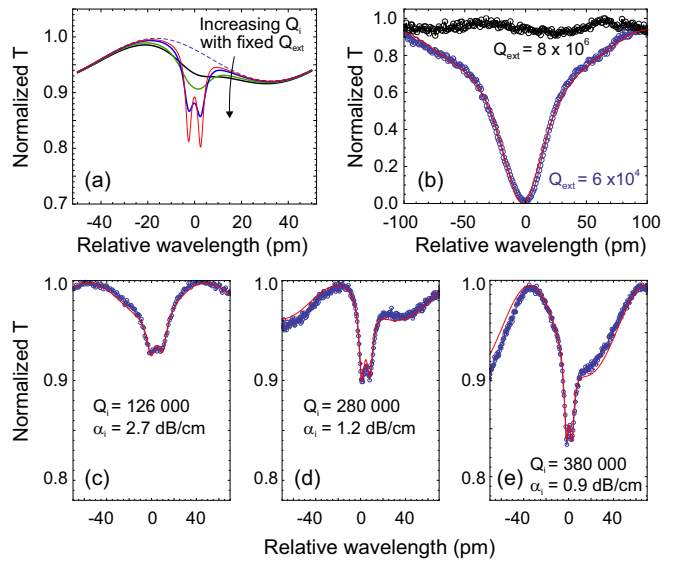}
\caption{ (a) Calculated spectral visibility of resonances of a 60~$\mu$m radius resonator with an external coupling $Q_{ext}=8 \times 10^6$ to the waveguide. The model takes into account the background Fabry-Peròt fringes due to waveguide-facets reflections. Modal splitting due to backscattering is also considered in order to evidence the effect of peak visibility change when the intrinsic loss of the resonator improves. (b) The peak visibility is near-zero in as-deposited samples with $Q_{ext}=8 \times 10^6$ (black dots), while a similar resonator is at critical coupling for a $Q_{ext}=6 \times 10^4$, revealing an intrinsic loss of about 5 dB/cm (blue dots and red fit-curve). An exposure to UV light cancels progressively nitride's net positive charge, which consequently decreases free-electrons concentration in the guiding Si layer. In conditions of fixed external coupling ($8 \times 10^6$), the resulting lower loss increases peak visibility. Example spectra (dots) and their fits (red line) are shown for (c) 5 h UV, (d) 23 h UV and (e) 23 h UV plus sintering in forming gas at 350$^\circ$C plus additional 2 h of UV. }
\label{fig:rings}
\end{figure}

Figure~\ref{fig:rings}(a) reports examples of spectra calculated following:
\begin{equation}
T(\omega)=\left|FP+\sqrt{\alpha_{e}/2}(m_s+m_c) \right|^2,
\label{eq:lor}
\end{equation}
where $FP$ is the Fabry-Peròt (FP) background of the bare waveguide transmission \cite{taebi2008modified}. Here, we have introduced doublet-resonances to account for formation of symmetric, $m_s$, and anti-symmetric, $m_c$, traveling waves within the resonator due to possible backscattering mechanisms \cite{borselli2005beyond}. These last are described as $m_{c,s}=-\frac{\sqrt{\alpha_e/2}}{\imath (\Delta\omega\pm\beta/2)-(\alpha_i+\alpha_e)/2}$, where $\Delta\omega$ is the frequency detuning from resonance and $\beta$ is the loss associated with the backscattering (fixed in these examples). Equation \ref{eq:lor} has been considered since this is the situation of our experiments with rings when the UV improvement results into narrower resonances, as will be discussed in the following.

We notice that when $\alpha_i\ll \alpha_e$, no resonant features can be observed out from the oscillating FP background (dashed line, Fig.~\ref{fig:rings}(a)). A resonant dip starts to appear when $\alpha_i$ is decreased, first, appearing as an ``unstructured bump" and transforming progressively into a brighter and well defined doublet when $\alpha_i \lesssim \alpha_e$.

Experimentally, we have realized on the same chip nominally identical ring resonators (radius of 60 $\mu$m and width of 1300 nm) coupled to the waveguides through three different gaps of $g=$ 800 nm, 1620 nm and 1800 nm. In particular, the largest gap provides with an external coupling $Q_{e}\approx 8\times 10^6$. This situation was considered to fulfill a critical coupling, $Q_{e}\approx Q_i$, based on the numerical simulations for lowest possible intrinsic loss of resonators (bend-loss limited), without accounting for scattering losses. Figure~\ref{fig:rings}(b) shows examples of typical spectra for as-prepared devices at two different coupling gaps. Namely, in the case of $g=$1800~nm we observe a featureless spectrum of the typical waveguide transmission, which means that the resonator's intrinsic loss is much higher than that of the coupling. In other words, we deal with the strongly-undercoupled $Q_i\ll Q_e$ situation. In fact, for devices with $g=$800~nm we observe critically coupled resonances, from which we extract $Q_i=Q_e\approx 6\times 10^4$, corresponding to an intrinsic loss of about $\alpha_i\approx5$ dB/cm. 

Panels (c), (d) and (e) of Fig.~\ref{fig:rings} show the evolution of the rings spectrum upon exposure to UV at various times. All the results are for a coupling gap of 1800~nm. We note that while devices with an intermediate gap ( $g=1620$ nm) provide with deeper resonances, they hide important spectral features of doublets due to larger coupling loss ($Q_{e}\approx 5\times 10^5$). Therefore, we concentrate analyzing clear doublets which manifest in devices with highest $Q_e$. The spectral form of the doublets and their peak transmission are more sensible to the intrinsic loss value, therefore, they permit extraction of $\alpha_i$ with higher accuracy with respect to the Beer-Lambert method, described previously. We notice that a five-hour UV treatment improves the loss to 2.7 dB/cm (Fig.~\ref{fig:rings}(c)) and, further, down to $\alpha_i\approx 1.2$ dB/cm after 23 h (panel (d)).

Contrary to expectations from the results of capacitance measurements, we did not observe further improvements of losses upon exposing devices to UV for times longer than 21 h. Our conclusion, at this point is that in the case of devices we either deal with residual scattering losses due to fabrication or the SOI structure provides with other charge-related losses which cannot be improved with UV exposure. These last can originate from positive charges situated at the Si/BOX interface or in the BOX oxide itself. For this, we performed additional sintering of the chips and a UV-improved (23 h) MOS test sample at 350$^\circ$C for 3 hours in forming gas to improve the Si/SiO$_2$ interface. Control MOS capacitance measurements showed that some positive charge in the Si$_3$N$_4$ was re-activated, which rapidly annihilated upon a post-sintering exposure to UV for 2 h. We repeated the same procedure on the device chip and measured the rings spectra. Figure~\ref{fig:rings}(e) shows an example spectrum of resonances, where the loss has been further improved down to 0.9 dB/cm. These results suggest that a certain amount of charge was, in fact, present in the SOI device and was partially neutralized after the sintering procedure. 

Further UV exposures did not improve the situation, from which we conclude that the main part of remaining loss originates from scattering. We would like to draw the reader's attention to the following observation. Contrary to the unpatterned Si$_3$N$_4$ layer in the MOS test samples, the nitride film in our devices has a finite width, while the SOI layer is infinite in plane. Due to this, a quasi-2D electronic potential well is created within the SOI and underneath the strip, therefore, it is possible that residual positive interface charges bend down silicon's electronic bands more than it occurs in nitride-free p-type Si regions. This may favor electrons to drift freely into the lower potential and contribute to minute free-carrier absorption. A dedicated study of electronic properties of the system may shed light on the exact situation, however, it is out of the scope of current work and has less impact on presented results.

\begin{figure}[t!]
	\centering
\includegraphics[width=0.95\columnwidth]{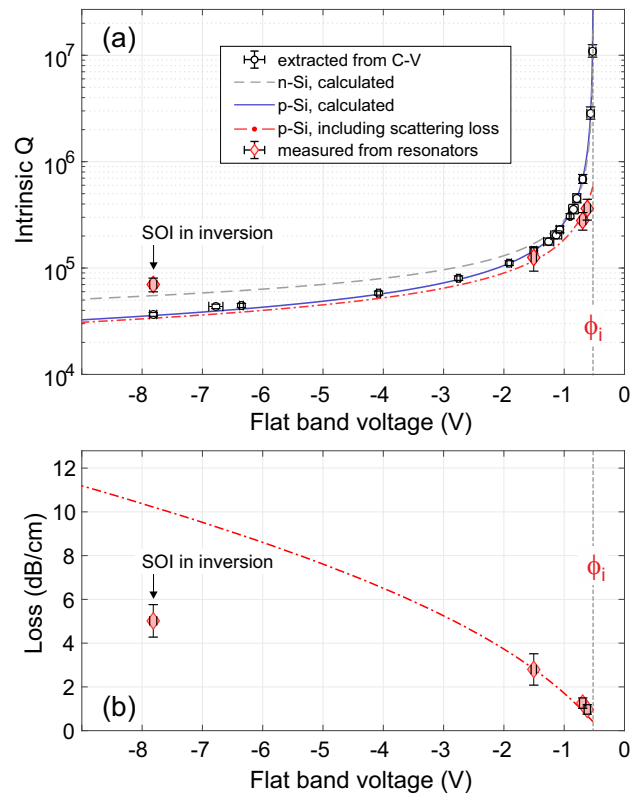}
\caption{ (a) The free-carrier-related intrinsic Q as a function of the UV-modified flat band voltage for p- and n-type Si, blue-continuous and grey-dashed curves, respectively. Quality factors, estimated from C-V measurements, are plotted as empty circles. The results from ring resonators are shown as diamonds ($\diamond$), with vertical error bars indicating the statistical error over a large number of analyzed resonances. The red ($-\cdot-$) curve is a fit to the three rightmost data points by considering an additional scattering $Q_{sc}$ of $6\times 10^5$. (b) The intrinsic loss $\alpha_i$, corresponding to extracted Q's from rings ($\diamond$), dash-dotted line is the calculated one considering the sacttering loss. }
\label{fig:losses}
\end{figure}

In Fig.~\ref{fig:losses}(a) we report the calculated trend of the intrinsic $Q_i$ against the flat-band voltage, $V_{fb}$ (continuous blue line). The variation of this last reflects the effect of UV exposure, and, thus, the gradual neutralization of the positive charge in the Si$_3$N$_4$ layer. The corresponding $Q_i$ is then related to $V_{fb}$ via
\begin{subequations}
\begin{equation}
Q_i=\frac{2\pi n_g}{\alpha_i(V_{fb}) \lambda}=\frac{2 \pi n_g}{\lambda}\times 0.939 \Delta P(V_{fb})^{-1.085},
\label{eq:Qint}
\end{equation}
\begin{equation}
\Delta P(V_{fb})=\frac{|C_{max} (V_{fb}-\phi_i)|}{q d_{Si} A},
\label{eq:DeltaP}
\end{equation}
\end{subequations}
where $n_g$ is the group index of the resonator mode, $q$ is the elementary charge, $d_{Si}$ is the thickness of the SOI layer, $A$ is the area of the gate contanct, and $\Delta P$ is the concentration of free carriers related to the optical loss $\alpha_i$ following empirical estimations from Ref.~\cite{degallaix2013bulk} for p-type Si. 

For comparison, we plot as open circles ($\circ$) the results estimated from experimental C-V curves and as diamonds ($\diamond$) the results from measurements on ring resonators. The vertical error bars in the first case come from at least five different positions per sample, while for rings we consider the standard deviations of measured $Q$'s over at least ten different resonances per ring. We notice, that while C-V results show a good match with the calculations, the data from rings deviate both for the as-prepared case as well as after UV treatments. The explanation of these observations is as follows: 
\begin{itemize}
\item [(i)] according to the capacitance measurements, the guiding SOI layer in the as-prepared samples is in conditions of complete inversion of the conductivity type. Thus, the measured $Q$'s find a better accordance with the theoretical curve if we consider that the SOI layer is an ``effective" n-type silicon \cite{degallaix2013bulk} (dashed grey curve).
\item [(ii)] the rest of data can be fit better if we consider a limiting residual scattering loss with an associated $Q_{sc}\approx 6\times10^5$ (red dash-dot curve).  
\end{itemize}

Finally, in Fig.~\ref{fig:losses}(b) we show the corresponding loss values from resonator's data and the calculated curve for  p-type Si in the situation of residual scattering loss. These results show the advantage of loss-estimation method from resonant features with respect to a classical Beer-Lambert approach. The estimated residual loss $\alpha_{sc}\approx0.55$ dB/cm, which we attributed to scattering, perhaps, may still contain a component associated to charging effects at the Si/SiO$_2$ interface or within the bulk of the BOX oxide. At present, this aspect is uncertain and is a subject of further investigations.

\section{Conclusions}
We reported in this work the design, fabrication and characterization of high-Q micro-optical components on an ultra-thin SOI platform. Waveguiding is supported by loading a 27~nm-thick SOI slab layer with micron-wide stripes of stoichiometric Si$_3$N$_4$. Such a configuration omits the need to etch physical boundaries in the SOI layer and, therefore, foresees to suppress largely the scattering loss \citep{stefan2015ultra}. 

Contrary to expectations, we revealed that the as-prepared devices were subject to significant loss of about 5 dB/cm. We related this to free-carrier absorption effects due to the presence of a large amount of electrical charge in the Si$_3$N$_4$ layer, originating from paramagnetic $^+$Si$\equiv$N$_3$ defects (dangling bonds). This hypothesis was, in a first step, confirmed by detailed MOS capacitance measurements, revealing a complete inversion of the conductivity type of the p-type Si substrate. As a result, the resistivity of the original substrate was reduced from $15~\Omega\cdot$cm down to $0.03~\Omega\cdot$cm, confirming the observed large optical losses. Next, we exposed test samples to 254~nm UV light for different times and observed gradual neutralization of the space charge in the nitride layer. Our estimations from these experiments showed that the equivalent areal electrical charge can be reduced by three orders of magnitude reaching a value of $\sigma\approx3 \times 10^{9}$ cm$^{-2}$, comparable to the density of charge traps of very-high-quality oxide/Si interfaces  \cite{zhang2006electronic}.  

Finally, UV-exposure was performed on SOI devices and the optical loss was directly measured from spiral waveguides by Beer-Lambert approach and ring resonators from the spectral linewidth of resonances. We revealed a net improvement of losses down to 0.9 dB/cm at longest exposures, improving the rings intrinsic Q-factors from 60,000 to 500,000. In addition, a 0.5~dB/cm contribution from other loss channels, such as residual scattering and possibly charging of the bottom cladding oxide, was also estimated.

We foresee that these results will go far beyond the target of the current study. The permanent nature of the UV-neutralization of the charges in Si$_3$N$_4$ can have important implications for the design of micro-photonic devices. For example, charged domains of material can be alternated to create static electrical poling on top of waveguiding components by using appropriate masking during UV-illumination. The sign and the amount of bulk charges are typical for various SiN$_x$'s deposited by different techniques, therefore, their modulation with UV light can be further studied in view of photonic applications. Last, but not least, it appears promising to implement the UV-exposure procedure in a currently hot topic such as the studies of the origin of dressed $\chi^2$ nonlinearities in nitride-strained silicon waveguides \cite{cazzanelli2012second,schriever2015second}. Finally, our results may open the door to feature developments of compactly integrated micro-photonic components with MOFSET electronics on the same ultra-thin SOI platform.

\section{ACKNOWLEDGEMENTS}
The authors gratefully thank Georg Pucker for support and fruitful discussions and Lorenzo Pavesi for providing access to optical measurement facility of NanoScience Laboratory at the University of Trento. The authors also acknowledge fabrication facility support by the Micro-Nano Fabrication Laboratory of FBK. Funding support is provided by the Italian Ministry of Education, Universities and Research under award number PRIN-NEMO 2015KEZNYM.

\bibliography{stripsoirefs} 

\end{document}